\documentclass[letterpaper,11pt]{article}

\usepackage{fullpage}
\usepackage{amsmath}
\usepackage{amsfonts}
\usepackage{amssymb}
\usepackage{amsthm}

\newcommand{\op}[1]{\operatorname{#1}}

\def\defeq {\stackrel{\mathrm{def}}{=}}

\newenvironment{mylist}[1]{\begin{list}{}{
    \setlength{\leftmargin}{#1}
    \setlength{\rightmargin}{0mm}
    \setlength{\labelsep}{2mm}
    \setlength{\labelwidth}{8mm}
    \setlength{\itemsep}{0mm}}}
    {\end{list}}

\newcommand{\tinyspace}{\mspace{1mu}}

\newcommand{\bra}[1]{\langle #1|}
\newcommand{\ket}[1]{|#1\rangle}

\newcommand{\tensor}{\otimes}
\newcommand{\norm}[1]{\left\lVert\tinyspace#1\tinyspace\right\rVert}
\newcommand{\tnorm}[1]{\norm{#1}_{\mathrm{tr}}}
\newcommand{\dnorm}[1]{\norm{#1}_{\diamond}}
\newcommand{\abs}[1]{\left\lvert\tinyspace #1 \tinyspace\right\rvert}

\newcommand{\fidelity}[2]{F(#1, #2)}

\newcommand{\density}[1]{\mathbf{D}(\mathcal{#1})}
\newcommand{\unitary}[1]{\mathbf{U}(\mathcal{#1})}

\newcommand{\transform}[1]{\mathbf{T}(\mathcal{#1})}
\newcommand{\identity}[1]{I_{\mathcal{#1}}}

\newcommand{\tr}{\operatorname{tr}}
\newcommand{\ptr}[1]{\tr_\mathcal{#1}}
\newcommand{\ceil}[1]{\left\lceil #1 \right\rceil}
\newcommand{\floor}[1]{\left\lfloor #1 \right\rfloor}
\newcommand{\prob}[1]{\textup{\textsf{#1}}}
\newcommand{\class}[1]{\textup{\textrm{#1}}}

\newtheorem{theorem}{Theorem}[section]
\newtheorem{lemma}[theorem]{Lemma}
\newtheorem{prop}[theorem]{Proposition}

\theoremstyle{definition}
\newtheorem{defn}[theorem]{Definition}
\newtheorem{proto}[theorem]{Protocol}
\newtheorem*{problem}{Problem}

\begin{document}

  
\title{\bf\Large
    On the hardness of distinguishing mixed-state\\ quantum computations
}

\author{
  Bill Rosgen\\
  Department of Computing Science\\
  University of Alberta\\
  Edmonton, Alberta, Canada
  \and
  John Watrous\\
  Department of Computer Science\\
  University of Calgary\\
  Calgary, Alberta, Canada
}

\date{July 22, 2004}

\maketitle

\begin{abstract}
  This paper considers the following problem.
  Two mixed-state quantum circuits $Q_0$ and $Q_1$ are given, and the goal
  is to determine which of two possibilities holds:
  (i) $Q_0$ and $Q_1$ act nearly identically on all possible quantum state
  inputs, or
  (ii) there exists some input state $\rho$ that $Q_0$ and $Q_1$ transform
  into almost perfectly distinguishable outputs.
  This problem may be viewed as an abstraction of the following problem: given
  two physical processes described by sequences of local interactions, are the
  processes effectively the same or are they different?
  We prove that this problem is a complete promise problem for the class
  \class{QIP} of problems having quantum interactive proof systems, and is
  therefore \class{PSPACE}-hard.
  This is in sharp contrast to the fact that the analogous problem for
  classical (probabilistic) circuits is in \class{AM}, and for unitary
  quantum circuits is in \class{QMA}.
\end{abstract}

\section{Introduction}

Randomness is a fundamental concept in complexity theory and cryptography
that is sometimes under-emphasized in the study quantum computing.
For example, the most typically used quantum computational model is the
unitary quantum circuit model restricted to pure quantum states; and
although this model can simulate randomized computations, in some sense
there is really no randomness at all in a unitary circuit computation.
Indeed, in the framework of quantum information, pure states and unitary
computations may be viewed as being analogous to definite logical states
and deterministic computations, with more general types of states and
non-unitary operations being possible.
In particular, quantum states may be mixed as opposed to pure,
arising for example when a probability distribution over pure states
is considered, and operations such as measurements and noise may be
non-unitary but physically possible.

A variant of the quantum circuit model allowing mixed states and non-unitary
operations was introduced by Aharonov, Kitaev, and Nisan~\cite{AharonovK+98}.
They showed that this more general model is in fact equivalent in
power to the unitary quantum circuit model.
The principle behind this equivalence is the fact that arbitrary physically
realizable quantum operations, including irreversible deterministic
computations, random coin-flips, measurements, noise, and so on, can be
described by unitary operations acting on larger systems.

However, while the two quantum circuit models are equivalent in computational
power, it is a misconception that they are identical, and that 
there is no loss of generality in restricting ones attention to fully
reversible quantum computational models.
Indeed, in some restricted settings the equivalence of the models breaks down.
For instance, it is not known if unitary quantum computations can simulate
classical randomized computations in bounded space.
For quantum finite automata the situation is much more alarming.
Here, unitarity imposes a restriction that provably weakens the model over
the usual deterministic (but irreversible) model; and while a definition based
on mixed-states gives a natural and more satisfying model that generalizes
classical (deterministic and probabilistic) finite automata, the weaker and
less motivated unitary model has received far more attention.

In this paper we describe a different sense in which the mixed-state quantum
circuit model differs significantly from the unitary quantum circuit model.
Our interest is with the computational complexity of problems about
quantum circuits, and in particular our focus is on the following problem.
Assume two mixed-state quantum circuits $Q_0$ and $Q_1$, which agree on the
number of input qubits and on the number of output qubits, are given.
For any input state $\rho$, let $Q_0(\rho)$ and $Q_1(\rho)$ denote the mixed
states obtained by running $Q_0$ and $Q_1$, respectively, on input $\rho$.
It is promised that either (i)~$Q_0(\rho)$ and $Q_1(\rho)$ are almost identical
for all states $\rho$, or (ii)~there exists an input state $\rho$ for which
$Q_0(\rho)$ and $Q_1(\rho)$ are very different, and the goal is to determine
which of these possibilities holds.
(A natural way to formalize the notions of $Q_0(\rho)$ and $Q_1(\rho)$ being
``almost identical'' and ``very different'' is discussed in the next section.)
This problem is phrased as a promise problem because it would be
artificially difficult if it were necessary to distinguish cases when the
distances between $Q_0(\rho)$ and $Q_1(\rho)$ are close to some threshold.
Even with such a promise, however, we show that this problem is
\class{PSPACE}-hard.
More specifically, we show that this problem is a complete promise
problem for the class \class{QIP} of problems possessing quantum
interactive proof systems.
In contrast, the classical analogue of this problem, to distinguish between
two probabilistic boolean circuits, is easily shown to be contained in the
class \class{AM}, while the variant of the problem where $Q_0$ and $Q_1$ are
unitary quantum circuits is contained in \class{QMA} \cite{JanzingW+03}.
According to our current state of knowledge this represents a significant
difference in hardness, given that $\class{AM} = \class{PSPACE}$ and
$\class{QMA} = \class{PSPACE}$ both seem unlikely.

It is natural to attribute the apparent difference in hardness of the above
problems to the presence of both randomness and quantum computation in the
mixed-state quantum circuits variant of the problem---removing either
randomness (leaving a unitary model) or quantum computation (leaving a
classical probabilistic model) results in a reduction in complexity.
This example underscores the distinction between unitary and mixed-state
quantum models.

The above problem is also interesting for the much different reason that it
abstracts the following natural physical problem: given two physical
processes, are they effectively the same or are they different?
Under the assumption that the physical processes in question are described in
terms of local interactions among particles that can implement qubits and
simulate mixed-state quantum computations, it follows that even to solve this
problem approximately is \class{PSPACE}-hard.

Finally, we are hopeful that the completeness of the problem discussed in
this paper may lead to new results on the structural properties of the
class \class{QIP}.
For example, it is currently known that
$\class{PSPACE}\subseteq\class{QIP}\subseteq\class{EXP}$ \cite{KitaevW00},
but no strong evidence has yet been provided that suggests either containment
should be an equality or a proper containment.

The rest of this paper is organized as follows.
In Section~\ref{sec:preliminaries} we discuss relevant background
on mixed-state quantum circuits and other aspects of quantum information,
and in Section~\ref{sec:problem} we state and discuss the definition of the
computational problem of distinguishing mixed-state quantum circuits
being considered.
The main hardness result is proved in Section~\ref{sec:hardness}.
We conclude with Section~\ref{sec:conclusion}, which mentions some
open questions relating to the topic of the paper.


\section{Preliminaries}
\label{sec:preliminaries}


\subsection{Admissible operations and mixed-state quantum circuits}
\label{sec:circuits}

We begin by discussing admissible quantum operations together with the
mixed-state quantum circuit model of Aharonov, Kitaev, and
Nisan~\cite{AharonovK+98}.

For positive integers $k$ and $l$, consider the set of operations mapping
$k$-qubit states to $l$-qubit states that correspond to physically
possible operations (in an idealized sense).
Quantum information theory gives a simple description of this set of
operations, sometimes called the set of {\em admissible operations}.
Specifically, an operation $\Phi$ from $k$ qubits to $l$ qubits is admissible
if its action on density matrices is linear, trace-preserving, and completely
positive.
This means that if $\rho$ is a density matrix on $k + m$ qubits for some
arbitrary value of $m$, and $\Phi$ is performed on the first
$k$ qubits of $\rho$, then the result is a valid density matrix on
$l + m$ qubits.
In symbols, $(\Phi\otimes I_m)(\rho)$ is a density matrix, where $I_m$ denotes
the identity mapping on $m$ qubit states.
Examples of admissible operations include unitary operations (which
require that $k = l$), irreversible classical computations from $k$ bits to
$l$ bits, and the operations of adding qubits in some specified state and
discarding qubits.

A quantum gate of type $(k,l)$ is a gate that takes $k$ qubits as input and
outputs $l$ qubits, and corresponds to some admissible operation.
Mixed-state quantum circuits are circuits that consist of some finite
collection of such gates along with acyclic input/output relations
among these gates.
A given mixed-state quantum circuit will have some number $n$ of input qubits
and some number $m$ of output qubits.
Using the same terminology for circuits as for gates, we may say that a
circuit is of type $(n,m)$ when this is the case, and more generally
we say that an operation is of type $(n,m)$ if it maps $n$ qubit states to
$m$ qubit states.
Thus, a circuit $Q$ of type $(n,m)$ specifies some admissible operation of type
$(n,m)$, and when convenient we also let $Q$ denote this admissible operation.

\begin{figure}
\begin{center}
\setlength{\unitlength}{1400sp}%
\begingroup\makeatletter\ifx\SetFigFont\undefined%
\gdef\SetFigFont#1#2#3#4#5{%
  \reset@font\fontsize{#1}{#2pt}%
  \fontfamily{#3}\fontseries{#4}\fontshape{#5}%
  \selectfont}%
\fi\endgroup%
\begin{picture}(6024,5424)(2389,-5773)
\thinlines
\put(4200,-5761){\framebox(2400,5400){$U$}}
\put(1000,-1711){\line( 1, 0){3200}}
\put(1000,-1861){\line( 1, 0){3200}}
\put(1000,-2011){\line( 1, 0){3200}}
\put(1000,-661){\line( 1, 0){3200}}
\put(1000,-961){\line( 1, 0){3200}}
\put(1000,-1261){\line( 1, 0){3200}}
\put(1000,-1561){\line( 1, 0){3200}}
\put(1000,-811){\line( 1, 0){3200}}
\put(1000,-1411){\line( 1, 0){3200}}
\put(1000,-1111){\line( 1, 0){3200}}
\put(6600,-1711){\line( 1, 0){3200}}
\put(6600,-661){\line( 1, 0){3200}}
\put(6600,-961){\line( 1, 0){3200}}
\put(6600,-1261){\line( 1, 0){3200}}
\put(6600,-1561){\line( 1, 0){3200}}
\put(6600,-811){\line( 1, 0){3200}}
\put(6600,-1411){\line( 1, 0){3200}}
\put(6600,-1111){\line( 1, 0){3200}}
\put(3000,-3361){\line( 1, 0){1200}}
\put(3000,-3211){\line( 1, 0){1200}}
\put(3000,-3511){\line( 1, 0){1200}}
\put(3000,-3811){\line( 1, 0){1200}}
\put(3000,-4111){\line( 1, 0){1200}}
\put(3000,-4411){\line( 1, 0){1200}}
\put(3000,-3661){\line( 1, 0){1200}}
\put(3000,-4261){\line( 1, 0){1200}}
\put(3000,-3961){\line( 1, 0){1200}}
\put(3000,-4561){\line( 1, 0){1200}}
\put(3000,-4861){\line( 1, 0){1200}}
\put(3000,-5161){\line( 1, 0){1200}}
\put(3000,-5461){\line( 1, 0){1200}}
\put(3000,-4711){\line( 1, 0){1200}}
\put(3000,-5311){\line( 1, 0){1200}}
\put(3000,-5011){\line( 1, 0){1200}}
\put(6600,-3361){\line( 1, 0){1200}}
\put(6600,-3211){\line( 1, 0){1200}}
\put(6600,-3061){\line( 1, 0){1200}}
\put(6600,-2911){\line( 1, 0){1200}}
\put(6600,-2761){\line( 1, 0){1200}}
\put(6600,-2611){\line( 1, 0){1200}}
\put(6600,-3511){\line( 1, 0){1200}}
\put(6600,-3811){\line( 1, 0){1200}}
\put(6600,-4111){\line( 1, 0){1200}}
\put(6600,-4411){\line( 1, 0){1200}}
\put(6600,-3661){\line( 1, 0){1200}}
\put(6600,-4261){\line( 1, 0){1200}}
\put(6600,-3961){\line( 1, 0){1200}}
\put(6600,-4561){\line( 1, 0){1200}}
\put(6600,-4861){\line( 1, 0){1200}}
\put(6600,-5161){\line( 1, 0){1200}}
\put(6600,-5461){\line( 1, 0){1200}}
\put(6600,-4711){\line( 1, 0){1200}}
\put(6600,-5311){\line( 1, 0){1200}}
\put(6600,-5011){\line( 1, 0){1200}}
\put(6600,-2461){\line( 1, 0){1200}}
\put(1000,-2161){\line( 1, 0){3200}}
\put(1000,-2311){\line( 1, 0){3200}}
\put(6600,-2311){\line( 1, 0){1200}}
\put(1000,-2461){\line( 1, 0){3200}}
\put(1000,-2611){\line( 1, 0){3200}}
\put(1400,-5700){\framebox(1600,2700){$\ket{0^{2l}}$}}
\put(7800,-5700){\framebox(1800,3600)[t]{\parbox{1.3cm}{\vspace{9mm}
	\begin{center}\small traced\\[-1mm] out\end{center}}}}
\put(0,-1775){$\rho\left\{\rule[-6mm]{0mm}{0mm}\right.$}
\put(9800,-1350){$\left.\rule[-4mm]{0mm}{0mm}\right\}\Phi(\rho)$}
\end{picture}
\end{center}
\caption{The unitary operation $U$ simulates the admissible operation $\Phi$.}
\label{fig:unitary_simulation}
\end{figure}
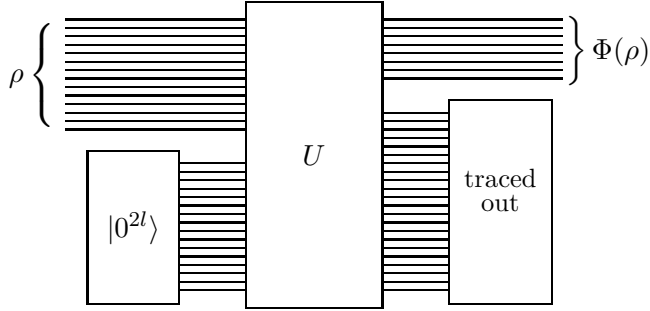

A necessary and sufficient condition for an operation $\Phi$ of type $(k,l)$
to be admissible is that there exists a unitary operation $U$ acting on
$k + 2l$ qubits such that the following holds.
If the first $k$ qubits are set to state $\rho$ and the remaining $2l$ qubits
are initialized to the $\ket{0}$ state, the operation $U$ is
applied, and finally the last $k+l$ qubits are discarded (or
{\em traced out}), the resulting state on the remaining $l$ qubits is
$\Phi(\rho)$.
This situation is illustrated in Figure~\ref{fig:unitary_simulation}.
This fact is generally attributed to Choi~\cite{Choi75},
and a proof may be found in Kitaev, Shen, and Vyalyi~\cite{KitaevS+02}.
This process may be applied to each gate in a given circuit $Q$, resulting
in a unitary circuit $P$ that simulates $Q$ in a sense similar to the
situation pictured in Figure~\ref{fig:unitary_simulation}.
Under the assumption that each gate is of constant size, the number of
additional qubits required is linear in the number of gates of $Q$.

For the remainder of this paper it will be assumed that all mixed-state quantum
circuits under consideration are composed of gates from some reasonable finite
set.
In order to avoid a discussion of what exactly is meant by ``reasonable'', let
us for simplicity say that this means that if the gates are expressed as linear
mappings, then these mappings can be written as matrices consisting of
efficiently approximable numbers.
The point is to disallow difficult to compute information from being
somehow incorporated into the action of gates acting on a finite number
of qubits.
Assuming that such a finite set of quantum gates has been fixed, a quantum
circuit may easily be described classically.
It will not be necessary to discuss a particular method of encoding quantum
circuits beyond stating the assumption that the encoding is efficient,
reasonable, and disallows compact descriptions of large circuits.
Given such a classical description of a circuit $Q$, it is possible to
compute in polynomial time a description of a unitary quantum circuit $P$ that
simulates $Q$ in the sense described above.

A few additional requirements on the set of gates of which
mixed-state quantum circuits may be composed is required for the
hardness results proved in this paper. 
The requirements are that (i) the set of gates is universal for quantum
computation, meaning that any constant-size unitary operation can be
efficiently approximated by circuits composed of these gates, (ii) the gates
include a gate of type $(0,1)$ that introduces a qubit initialized
to the state $\ket{0}$, and (iii) the gates include the unique
gate of type $(1,0)$ that corresponds to discarding a qubit.


\subsection{Distance measures for quantum states and admissible operations}
\label{sec:distance}

The problem of distinguishing quantum circuits on which this paper focuses
requires a notion of distance between admissible operations.
The notion we will use, and which we claim is the most natural with respect
to the problem, is given by a norm known as the diamond norm.

Before discussing the diamond norm, we need to mention the trace norm,
which induces a distance measure between density matrices that is
analogous to the distance between probability distributions induced
by the 1-norm.
For a given square matrix $X$, the trace norm of $X$, denoted
$\|X\|_{\mathrm{tr}}$, is defined to be the sum of the singular values of $X$.
In case $X$ is Hermitian, $\tnorm{X}$ is also equal to the sum of the absolute
value of the eigenvalues of $X$.
Equivalent expressions for the trace norm (for general $X$) include
$\|X\|_{\mathrm{tr}} = \op{tr}\sqrt{X^{\dagger}X}$
and $\|X\|_{\mathrm{tr}} = \max\{|\op{tr}(XU)|\}$,
where the maximum is over all unitary $U$ having the same dimensions as $X$.

The quantity $\tnorm{\rho_0 - \rho_1}$ for given density matrices $\rho_0$
and $\rho_1$ has the following operational interpretation.
Given any binary-valued measurement, let us say that the measurement
is correct in the event that, on input $\rho_b$, the outcome of the
measurement is $b$, and is incorrect when the outcome is $\neg b$.
Assuming $\rho_0$ and $\rho_1$ are each given with probability 1/2,
the quantity $\tnorm{\rho_0 - \rho_1}/2$ represents the maximum over
all possible measurements
that the measurement is correct minus the probability
the measurement is incorrect.
Thus, $\tnorm{\rho_0 - \rho_1} = 2$ implies that $\rho_0$ and $\rho_1$ are
perfectly distinguishable by some measurement, while
$\tnorm{\rho_0 - \rho_1} = 1$, for example, implies that the maximum
probability of correctness for any measurement given $\rho_0$ and
$\rho_1$ uniformly is 3/4.
Obviously $\tnorm{\rho_0 - \rho_1} = 0$ implies $\rho_0 = \rho_1$,
and so no measurement can do better than random guessing in this case.

The trace norm may be extended to differences in admissible operations
in the following standard way:
if $\Phi$ and $\Psi$ are admissible, then
\[
\tnorm{\Phi-\Psi} \defeq \max\{\tnorm{\Phi(X) - \Psi(X)}\,:\,\tnorm{X} = 1\}.
\]
Unfortunately this norm has some unusual properties that make it unsuitable
for describing distances between admissible operations.
One problem is that the maximum may not be achieved when $X$ is
a density matrix, and another is that the value of the norm may change
if $\Phi$ and $\Psi$ are tensored with the identity operation on some
number of qubits.

With this in mind, one defines the {\em diamond norm} of the difference
$\Phi - \Psi$, for $\Phi$ and $\Psi$ admissible operations of type $(n,m)$,
as follows:
\[
\dnorm{\Phi - \Psi} \defeq \tnorm{\Phi \otimes I_n - \Psi \otimes I_n} =
\max\{\tnorm{(\Phi\otimes I_n)(X) - (\Psi\otimes I_n)(X)}\,:\,\tnorm{X} = 1\}.
\]
Here, $I_n$ denotes the identity operation on states of $n$ qubits and
the maximum is over all $2^{2n}\times 2^{2n}$ matrices $X$ (with
$\tnorm{X} = 1$).
The diamond norm was first defined and studied by Kitaev~\cite{Kitaev97}.
Further information on it may be found in Refs.~\cite{KitaevS+02}
and \cite{AharonovK+98}.
The maximum in the above definition always occurs for $X$ a density matrix
(and therefore for $X = \ket{\psi}\bra{\psi}$ for some unit vector
$\ket{\psi}$ by a simple convexity argument), and the quantity does not grow
if the identity is taken on more than $n$ qubits.
The second fact was already known but the first is new.
A more technical discussion of these facts can be found below in
Section~\ref{sec:technical-background}.

The diamond norm gives a similar characterization of the distinguishability
of admissible operations that the trace norm gives for states.
Specifically, the diamond norm of the difference between two admissible
operations characterizes the probability that the output of these two
operations can be distinguished, given that an input to the two operations
is chosen that maximizes the distinguishability of the outputs.
It is important to note that this includes the possibility that the
input is a state of a larger system on which the operations act on only
part.

Another useful way to measure the similarity between density
matrices is given by the fidelity. Specifically, the fidelity
between density matrices $\rho$ and $\xi$ is defined as:
\[
F(\rho,\xi)\:\defeq\:\op{tr}\sqrt{\sqrt{\rho}\,\xi\sqrt{\rho}}.
\]
The fidelity is a measure of similarity that is related to but different from
the trace norm.
Generally speaking, when two states are close together they have large fidelity
and small trace norm, and when far apart have small fidelity and large trace
norm.
At first glance the fidelity appears to be an unusual and possibly difficult
to use quantity, but in actuality it is often easier to use than the trace
norm.
(For instance, it is multiplicative with respect to tensor products.)
For all density matrices $\rho$ and $\xi$, it holds that
\[
1 - \frac{1}{2} \tnorm{\rho - \xi} \leq \fidelity{\rho}{\xi} \leq
\sqrt{1 - \frac{1}{4} \tnorm{\rho - \xi}^2}.
\]


\subsection{Quantum interactive proof systems}
\label{sec:qips}

Quantum interactive proof systems are interactive proof systems in which the
prover and verifier may exchange and process quantum information
\cite{KitaevW00, Watrous03-pspace}.
The class of problems having quantum interactive proof systems is denoted
\class{QIP} and is known to satisfy
$\class{PSPACE}\subseteq\class{QIP}\subseteq\class{EXP}$.

The main result of this paper, stated more formally in the next section,
establishes that the problem of distinguishing mixed-state quantum
circuits is \class{QIP}-complete.
This will be proved by first noting that a fairly straightforward
quantum interactive proof system exists for the problem, and second
by reducing a problem that was already known to be complete for \class{QIP}
to the circuit distinguishing problem.
In fact, the problem we will use for the reduction was only implicitly
proved to be complete for \class{QIP} in Ref.~\cite{KitaevW00}, but all
of the pieces needed to establish this fact are present in that paper.
The problem is as follows.

\begin{problem}[Close Images]
This problem is parameterized by constants $a,b\in[0,1]$ with $b < a$.
For such constants, define the promise problem $\prob{CI}_{a,b}$ as follows:
\vspace{2mm}

\noindent
\begin{tabular}{@{}lp{5.75in}@{}}
\textbf{Input:} &
    Mixed state quantum circuits $(Q_0, Q_1)$ of type $(n,m)$.\\[2mm]
\textbf{Yes:} &
    There exist $n$ qubit states $\rho_0$ and $\rho_1$ such that
    $F(Q_0(\rho_0),Q_1(\rho_1)) \geq a$.\\[2mm]
\textbf{No:} &
    For all $n$ qubit states $\rho_0$ and $\rho_1$,
    $F(Q_0(\rho_0),Q_1(\rho_1)) \leq b$.
\end{tabular}
\end{problem}

\noindent
The ``yes'' instances of the problem are therefore circuits whose images
are close with respect to fidelity, while the ``no'' instances are circuits
whose images are far apart.
Completeness of this promise problem for \class{QIP} holds for
any constants $a,b$ with $0<b<a\leq 1$.


\subsection{More notation and technical facts concerning distance measures}
\label{sec:technical-background}

The proofs in the sections that follow will require more precise notation
than has been necessary thus far, as well as a few key facts about the
distance measures discussed previously.
It is convenient to include these things at this point, but the reader
uninterested in the technical details of the proofs may safely skip the
remainder of this section.
For the most part our notation is standard and consistent with
Kitaev, Shen, and Vyalyi \cite{KitaevS+02}, which may be consulted for
further background information.

Hilbert spaces will be denoted by scripted letters, such as
$\mathcal{H}$, $\mathcal{K}$, etc.
It will always be the case in this paper that Hilbert spaces
have a standard orthonormal basis that is in correspondence with
binary strings of a given length.
We write, for instance, $\mathcal{H} = \mathcal{H}(\Sigma^n)$ when
the standard basis of $\mathcal{H}$ is in correspondence with
$\Sigma^n$, for $\Sigma = \{0,1\}$.
For given Hilbert spaces $\mathcal{H}$ and $\mathcal{K}$,
$\mathbf{L}(\mathcal{H},\mathcal{K})$ denotes the set of linear operators
from $\mathcal{H}$ to $\mathcal{K}$, and $\mathbf{L}(\mathcal{H})$ 
is shorthand for $\mathbf{L}(\mathcal{H},\mathcal{H})$.
The set $\mathbf{D}(\mathcal{H})$ consists of all positive semidefinite
operators on $\mathcal{H}$ having unit trace (i.e., all density matrices
over $\mathcal{H}$).
The set $\mathbf{U}(\mathcal{H},\mathcal{K})$  consists of all linear operators
from $\mathcal{H}$ to $\mathcal{K}$ that preserve the Euclidean norm.
Equivalently, $U^{\dagger}U = I_{\mathcal{H}}$ (the identity operator on
$\mathcal{H}$).
In case $\op{dim}(\mathcal{H}) = \op{dim}(\mathcal{K})$,
$\mathbf{U}(\mathcal{H},\mathcal{K})$ consists of those operators that are
unitary, and we write $\mathbf{U}(\mathcal{H})$ as a shorthand for
$\mathbf{U}(\mathcal{H},\mathcal{H})$.
The set $\mathbf{T}(\mathcal{H},\mathcal{K})$ consists of the linear operators
from $\mathbf{L}(\mathcal{H})$ to $\mathbf{L}(\mathcal{K})$.
Admissible operations are examples of such mappings, which in general will be
called transformations.

The partial trace is the admissible operation obtained by taking
the tensor product of the trace with the identity, and corresponds to
discarding part of a quantum system.
One writes $\mbox{tr}_{\mathcal{H}}$ to denote this operation when the trace
is on the space $\mathcal{H}$.
If $X\in \mathbf{L}(\mathcal{H})$ is positive semidefinite
and $\ket{\psi}\in\mathcal{H}\otimes\mathcal{K}$ satisfies
$\op{tr}_{\mathcal{K}}\ket{\psi}\bra{\psi} = X$, then $\ket{\psi}$ is said
to be a {\em purification} of $X$.
Such a purification always exists provided
$\op{dim}(\mathcal{K})\geq \op{rank}(X)$.
The fidelity has an alternate characterization in terms of purifications
that is important to a proof appearing later.
\begin{lemma}\label{lemma:fidelity-tracenorm}
  Let $\rho, \xi \in \density{H}$.
  Then for arbitrary purifications
  $\ket\psi, \ket\phi \in \mathcal{H \tensor K}$
  of $\rho$ and $\xi$, respectively, we have
  $\tnorm{ \ptr{H} \ket\psi \bra\phi } = \fidelity{\rho}{\xi}$.
  \begin{proof}
    Using one of the alternate characterizations of the trace-norm
    together with Uhlmann's Theorem and a well known fact about the
    unitary equivalence of purifications of a given state, we have
    \[
    \begin{aligned}
      \tnorm{ \ptr{H} \ket\psi \bra\phi }
      & = \max_{U \in \unitary{K}} \lvert
      \tr\left(\ptr{H}\ket\psi\bra\phi\right) U \rvert
      = \max_{U \in \unitary{K}} \lvert \tr \ket\psi \bra\phi
      (\identity{H} \tensor U)\rvert\\
      & = \max_{U \in \unitary{K}} \lvert \bra\phi (\identity{H} \tensor U)
      \ket\psi \rvert
      = F(\rho,\xi)
    \end{aligned}
    \]
    as claimed.
  \end{proof}
\end{lemma}

We now give a more general definition for the diamond norm, which is
consistent with the definition given previously for differences of admissible
transformations.

\begin{defn}
\label{def:diamond}
If $\Phi\in\mathbf{T}(\mathcal{H},\mathcal{K})$ then
\[
\dnorm{\Phi} \defeq \tnorm{\Phi\otimes I_{\mathbf{L}(\mathcal{G})}}
\]
where $\mathcal{G}$ is a Hilbert space with
$\op{dim}(\mathcal{G}) = \op{dim}(\mathcal{H})$.
\end{defn}

\noindent
It is known (see Ref.~\cite{KitaevS+02}) that increasing the dimension of
$\mathcal{G}$ gives no increase in
$\tnorm{\Phi\otimes I_{\mathbf{L}(\mathcal{G})}}$.

\begin{theorem}[Kitaev]
\label{theorem:diamnorm-tracenorm}
Let $\Phi \in \mathbf{T}(\mathcal{H},\mathcal{K})$, and let
$\mathcal{F}$ be a space of arbitrary finite dimension.
Then
\[
\tnorm{\Phi \otimes I_{\mathbf{L}(\mathcal{F})}}
\leq \dnorm{\Phi}.
\]
\end{theorem}

The following fact shows that the maximum in Definition~\ref{def:diamond}
occurs on a rank-one projection provided
$\Phi$ is the difference of two completely positive transformations.
In particular this holds when $\Phi$ is the difference of two admissible
operations.

\begin{lemma}  \label{lemma:max-for-dnorm}
  Let $\Phi\in\mathbf{T}(\mathcal{H},\mathcal{K})$ satisfy
  $\Phi = \Phi_0 - \Phi_1$ for $\Phi_0$ and $\Phi_1$ completely positive.
  Then there exists a Hilbert space $\mathcal{F}$ and a unit vector
  $\ket{\psi}\in\mathcal{H}\otimes\mathcal{F}$ such that
  \[
  \dnorm{\Phi} = \tnorm{(\Phi\otimes I_{\mathbf{L}(\mathcal{F})})(\ket{\psi}
    \bra{\psi})}.
  \]
  \begin{proof}
    Let $\mathcal{G}$ be a Hilbert space with
    $\dim\mathcal{G} = \dim\mathcal{H}$.
    Then
    \[
    \dnorm{\Phi} = \tnorm{\Phi \tensor I_{\mathbf{L}(\mathcal{G})}} =
    \max\{
    \tnorm{ (\Phi \tensor I_{\mathbf{L}(\mathcal{G})})(X)}\,
      :\,\tnorm{X} = 1 \}.
    \]
    Let $X\in\mathbf{L}(\mathcal{H}\otimes\mathcal{F})$ satisfy this maximum,
    let $\mathcal{A} = \mathcal{A}(\Sigma)$ be a Hilbert space corresponding
    to a single qubit, and let
    $Y\in\mathbf{L}(\mathcal{H}\otimes\mathcal{F}\otimes\mathcal{A})$
    be defined as
    \[
    Y = \frac{1}{2} X \tensor \ket{0}\bra{1}+\frac{1}{2}X^\dagger\tensor
    \ket{1}\bra{0}.
    \]
    We have $\tnorm{Y} = \tnorm{X} = 1$ and $Y = Y^{\dagger}$.
    
    The condition that $\Phi = \Phi_0 - \Phi_1$ for $\Phi_0$ and $\Phi_1$
    completely positive implies $\Phi(X)^{\dagger} = \Phi(X^{\dagger})$ for
    every $X\in\mathbf{L}(\mathcal{H})$.
    (In fact, the two conditions are equivalent.)
    Defining $\mathcal{F} = \mathcal{G}\otimes\mathcal{A}$, we therefore
    have that
    \begin{eqnarray*}
      \tnorm{(\Phi \tensor \identity{\mathbf{L}(\mathcal{F})})(Y)}
      & = &
      \frac{1}{2} \tnorm{ (\Phi\otimes I_{\mathbf{L}(\mathcal{G})})(X) \tensor
    \ket{0}\bra{1} + (\Phi\otimes I_{\mathbf{L}(\mathcal{G})})(X^\dagger)
    \tensor \ket{1}\bra{0}} \\
      & = &
      \frac{1}{2}\tnorm{ (\Phi\otimes I_{\mathbf{L}(\mathcal{G})})(X)}
      +\frac{1}{2}\tnorm{(\Phi\otimes I_{\mathbf{L}(\mathcal{G})})
	(X^\dagger)}\\
      & = &
      \frac{1}{2}\tnorm{ (\Phi\otimes I_{\mathbf{L}(\mathcal{G})})(X)}
      + \frac{1}{2}\tnorm{\left((\Phi\otimes I_{\mathbf{L}(\mathcal{G})})(X)
    \right)^\dagger}\\
      & = & \tnorm{(\Phi\otimes I_{\mathbf{L}(\mathcal{G})})(X)} \\
      & = & \dnorm{\Phi}.
    \end{eqnarray*}

    As $Y$ is Hermitian, we may write
    \[
    Y = \sum_i \lambda_i \ket{\psi_i}\bra{\psi_i},
    \]
    where $\{ \ket{\psi_i}\}$ is an orthonormal set of eigenvectors of
    $Y$ with real eigenvalues $\{\lambda_i\}$.
    As $\tnorm{Y} = 1$, we have $\sum_i |\lambda_i| = 1$.
    Now,
    \[
    \tnorm{(\Phi \tensor \identity{\mathbf{L}(\mathcal{F})})(Y)}
    \leq \sum_i \abs{ \lambda_i } \tnorm{(\Phi \tensor
      \identity{\mathbf{L}(\mathcal{F})})(\ket{\psi_i}\bra{\psi_i})},
    \]
    and because $\sum_i \abs{\lambda_i} = 1$ we have
    \[
    \tnorm{(\Phi \tensor \identity{\mathbf{L}(\mathcal{F})})
      (\ket{\psi_i}\bra{\psi_i})}
    \geq
    \dnorm{\Phi}
    \]
    for some $i$.
    Let $\ket{\psi} = \ket{\psi_i}$ for some value of $i$ for which this
    inequality is satisfied.
    Because $\tnorm{(\Phi \tensor \identity{\mathbf{L}(\mathcal{F})})
      (\ket{\psi}\bra{\psi})} \leq \dnorm{\Phi}$ by
    Theorem~\ref{theorem:diamnorm-tracenorm}, we have
    $\tnorm{(\Phi \tensor \identity{\mathbf{L}(\mathcal{F})})
      (\ket{\psi}\bra{\psi})} = \dnorm{\Phi}$
    as required.
  \end{proof}
\end{lemma}

\noindent
Strangely, this fact does not hold in general for the trace norm
$\tnorm{\Phi}$ in place of the diamond norm.


\section{The quantum circuit distinguishability problem}
\label{sec:problem}

The problem of distinguishing the actions of two circuits is an
interesting problem from a complexity theoretic standpoint.
The problem of distinguishing two classical circuits that do not
make use of randomness is in \class{NP}, as one can easily verify that two 
circuits have different outputs given an input on which they differ.
If the circuits use randomness they can be distinguished in \class{AM}
by a fairly straightforward protocol.
If we change the model to quantum circuits over pure states, which capture the
intuitive notion of deterministic computation using quantum information, the
complexity of the circuit distinguishability problem is in \class{QMA}
(which is essentially a quantum version of \class{NP}) as shown by
Janzing, Wocjan, and Beth~\cite{JanzingW+03}.
If we combine these models, moving to mixed state quantum circuits, where
non-unitary operations such as measurement can add randomness, we see what
appears to be a significant increase in the complexity of the problem.
The definition of the problem follows.

\begin{problem}[Quantum Circuit Distinguishability]
  The problem is parameterized by constants $a,b\in[0,2]$ with $b < a$.
  For such constants, define a promise problem $\prob{QCD}_{a,b}$ as follows:
  \vspace{2mm}

  \noindent
  \begin{tabular}{@{}lp{5.75in}@{}}
    \textbf{Input:} & Mixed-state quantum circuits $(Q_0, Q_1)$, both
    of the same type $(n,m)$.\\[2mm]
    \textbf{Yes:} & $\dnorm{Q_0 - Q_1} \geq a$\\[2mm]
    \textbf{No:} & $\dnorm{Q_0 - Q_1} \leq b$
  \end{tabular}
\end{problem}

\noindent
One may also consider the case where $a$ and $b$ are functions depending on
the input length, but this paper will focus on the case where $a$ and $b$
are constant.

At first glance this problem appears to be similar to the $\prob{CI}_{a,b}$
problem of the previous section, but we claim that the relation is not at
all obvious.
We feel the $\prob{QCD}$ problem is a more interesting problem, particularly
because it abstracts a natural physical problem and reveals an apparent
complexity-theoretic difference between pure and mixed state models as was
discussed previously.
In contrast, the \prob{CI} problem is really just a rephrasing, based on a
theorem in quantum information theory known as Uhlmann's Theorem, of the
problem that asks whether a given three-message quantum interactive proof
system can be made to accept with high probability.

We now observe that $\prob{QCD}_{a,b} \in \class{QIP}$ provided
$a$ and $b$ are constants with $b < a$.
(For variable $a$ and $b$, this fact holds if $a$ and $b$ are polynomial-time
computable and are separated by the reciprocal of some polynomial.)
A simple proof system for this problem is based on the ``blind taste-test''
idea that is frequently used in the study of interactive proofs.
Specifically, a prover attempting to prove that circuits $Q_0$ and $Q_1$
differ prepares a state $\rho$ on which they differ and sends the part
of $\rho$ on which the circuits act to the verifier.
The verifier applies either $Q_0$ or $Q_1$ randomly, sends the output
to the prover, and challenges the prover to identify which circuit was applied.

\begin{theorem}\label{theorem:QCD_in_QIP}
$\prob{QCD}_{a,b}\in\class{QIP}$
for any constants $a$ and $b$ with $0\leq b < a \leq 2$.
\end{theorem}

\noindent
The proof is based on the following protocol.

\begin{proto}[Quantum Circuit Distinguishability]\label{tnd_proto}
  Input to both $P$ and $V$ is $(Q_0, Q_1)$,
  where circuits $Q_0$ and $Q_1$ are assumed to both be of type $(n,m)$.
\begin{mylist}{5mm}
\item[1.]\label{step_send_rho}
  $V$ receives from $P$ an $n$-qubit quantum register $\mathsf{X}$.

\item[2.]\label{step_send_back}
  $V$ selects $i \in \{0,1\}$ uniformly and applies circuit $Q_i$ to
  $\mathsf{X}$.
  The result is an $m$-qubit register $\mathsf{Y}$, which $V$ sends to $P$.

\item[3.]\label{step_challenge}
  $V$ receives from $P$ some $j \in \{0,1\}$, and accepts if $i = j$,
  rejecting otherwise.
\end{mylist}
\end{proto}

\begin{proof}[Proof of Theorem~\ref{theorem:QCD_in_QIP}]
We will show that the verifier described in Protocol \ref{tnd_proto} admits a
quantum interactive
proof system for $\prob{QCD}_{a,b}$ with acceptance probability at least
$\frac{1}{2} + \frac{a}{4}$ on yes instances and acceptance probability
at most $\frac{1}{2} + \frac{b}{4}$ on no instances.
It suffices to prove that the maximum probability with which a prover
can cause the verifier described in Protocol~\ref{tnd_proto} to accept
is $\frac{1}{2} + \frac{1}{4}\dnorm{Q_0 - Q_1}$.

Let $\mathcal{H}$ be the Hilbert space corresponding to the
input qubits of $Q_0$ and $Q_1$, and let $\mathcal{K}$ be the Hilbert space
corresponding to the output qubits.
By Lemma~\ref{lemma:max-for-dnorm} there exists a Hilbert space
$\mathcal{G}$ and a unit vector $\ket{\psi}\in\mathcal{H}\otimes\mathcal{G}$
such that
$\dnorm{Q_0 - Q_1} =
\tnorm{(Q_0\otimes I_{\mathbf{L}(\mathcal{G})})(\ket{\psi}\bra{\psi})
-(Q_1\otimes I_{\mathbf{L}(\mathcal{G})})(\ket{\psi}\bra{\psi})}$.
Fix such a $\ket{\psi}$ and define
$\rho_0 = (Q_0\otimes I_{\mathbf{L}(\mathcal{G})})(\ket{\psi}\bra{\psi})$
and $\rho_1 = (Q_1\otimes I_{\mathbf{L}(\mathcal{G})})(\ket{\psi}\bra{\psi})$.
Now, let $\Pi_0$ and $\Pi_1 = I - \Pi_0$ be projection operators on
$\mathcal{K}\otimes\mathcal{G}$ that specify an optimal projective measurement
for distinguishing $\rho_0$ from $\rho_1$.
Such a measurement satisfies
$\op{tr}\Pi_0 (\rho_0 - \rho_1) =
\op{tr}\Pi_1 (\rho_1 - \rho_0) =
\frac{1}{2}\tnorm{\rho_0 - \rho_1}$.
Now, a strategy for the prover that convinces the verifier to accept with
probability $\frac{1}{2} + \frac{1}{4}\dnorm{Q_0 - Q_1}$ is as follows.
The prover prepares two registers $(\mathsf{X},\mathsf{Z})$ in state
$\ket{\psi}$ and sends $\mathsf{X}$ to the verifier.
Upon receiving $\mathsf{Y}$ from the verifier, the prover measures
$(\mathsf{Y},\mathsf{Z})$ with the measurement $\{\Pi_0,\Pi_1\}$ and
returns the result to the verifier.
It is a simple calculation to show that this measurement correctly determines
$i$ with probability
$\frac{1}{2} + \frac{1}{4}\tnorm{\rho_0 - \rho_1} =
\frac{1}{2} + \frac{1}{4}\dnorm{Q_0 - Q_1}$.

The probability of acceptance attained by the above prover strategy is
optimal, which may be argued as follows.
Let $\xi$ denote the mixed state of the register $\mathsf{X}$ together
with any private qubits of the prover, which we represent as a register
$\mathsf{Z}$, immediately after the first message is sent.
As before, we let $\mathcal{G}$ denote the Hilbert space corresponding
to the prover's private qubit register $\mathsf{Z}$.
The verifier applies either $Q_0$ or $Q_1$, which causes the pair
$(\mathsf{Y},\mathsf{Z})$ to be in state
$(Q_0\otimes I_{\mathbf{L}(\mathcal{G})})(\xi)$ with probability 1/2 and
$(Q_1\otimes I_{\mathbf{L}(\mathcal{G})})(\xi)$ with probability 1/2.
The register $\mathsf{Y}$ is sent to the prover.
The prover's final message to the verifier is measured by the verifier,
resulting in a single bit.
This process may be viewed as a binary valued measurement of registers
$(\mathsf{Y},\mathsf{Z})$.
The probability that this measurement is correct is bounded above by
\[
\frac{1}{2} + \frac{1}{4}\tnorm{
(Q_0\otimes I_{\mathbf{L}(\mathcal{G})})(\xi) -
(Q_1\otimes I_{\mathbf{L}(\mathcal{G})})(\xi)}
\leq \frac{1}{2} + \frac{1}{4}\dnorm{Q_0 - Q_1}
\]
as required.
\end{proof}

Note that a simple variant of the protocol described above gives an 
ordinary interactive proof system for the classical probabilistic version of
the Circuit Distinguishability problem.
As the proof system uses a constant number of messages, this demonstrates
that the classical variant of the problem is contained in \class{AM}.


\section{QIP-hardness of distinguishing quantum circuits}
\label{sec:hardness}

In this section we prove that $\prob{QCD}_{a,b}$ is hard, with respect to Karp
reductions, for the class \class{QIP} for any choice of constants $a$ and $b$
with $0<b<a<2$.

\begin{theorem}\label{theorem:main}
$\prob{QCD}_{2-\varepsilon,\varepsilon}$ is \class{QIP}-complete for every
$\varepsilon >0$.
\end{theorem}

\noindent
This theorem is proved in two stages.
First, the Close Images problem (for some appropriate choice of parameters) is
reduced to $\prob{QCD}_{1,1/4}$, implying \class{QIP}-hardness of
$\prob{QCD}_{1,1/4}$.
Then, it is argued that $\prob{QCD}_{1,1/4}$ reduces to
$\prob{QCD}_{2-\varepsilon,\varepsilon}$ for any constant $\varepsilon > 0$,
which is sufficient to establish the main result.
In fact, $\prob{QCD}_{2-\varepsilon,\varepsilon}$ remains \class{QIP}-hard
even when $\varepsilon$ is not constant, but rather is an exponentially small
function of the input size.


\subsection{Overview of proof}
\label{sec:proof-overview}

The input to the \prob{CI} problem is a description of two circuits $Q_0$ and
$Q_1$, both of type $(n,m)$ for nonnegative integers $n$ and $m$.
The reduction will transform the description of these two circuits into a
description of two circuits $(R_0,R_1)$ that form an input to the \prob{QCD}
problem.

As discussed in Section~\ref{sec:circuits} we may convert $Q_0$ and $Q_1$
into unitary circuits $P_0$ and $P_1$, acting on $n+k=m+l$ qubits, that
simulate $Q_0$ and $Q_1$.
Here, $k$ is the number of initialized qubits introduced into the circuit
and $l$ is the number of ``garbage'' qubits that are discarded at the end
of the simulation.
The assumption that $P_0$ and $P_1$ act on the same number of qubits can be
made without loss of generality, as additional dummy qubits could be added to
either circuit as necessary.
Given descriptions of $P_0$ and $P_1$ it is possible to efficiently
construct a unitary circuit $P$ that acts on one more qubit than $P_0$ and
$P_1$, and uses this additional qubit as a control to determine which of the
two circuits $P_0$ or $P_1$ to perform.
In other words, $P(\ket{0}\ket{\psi}) = \ket{0}P_0\ket{\psi}$ and
$P(\ket{1}\ket{\psi}) = \ket{1}P_1\ket{\psi}$ for any $\ket{\psi}$.

Next, define
$D(\sigma) = \ket{0}\bra{0}\sigma\ket{0}\bra{0} +
\ket{1}\bra{1}\sigma\ket{1}\bra{1}$.
This is an admissible operation on a single qubit that represents
the process known as decoherence.
Informally, the qubit is measured in the standard basis and the result is
forgotten.
If this gate is not included in the choice of basis gates, it can
easily be constructed from gates in any basis satisfying the
requirements discussed in Section~\ref{sec:circuits}.

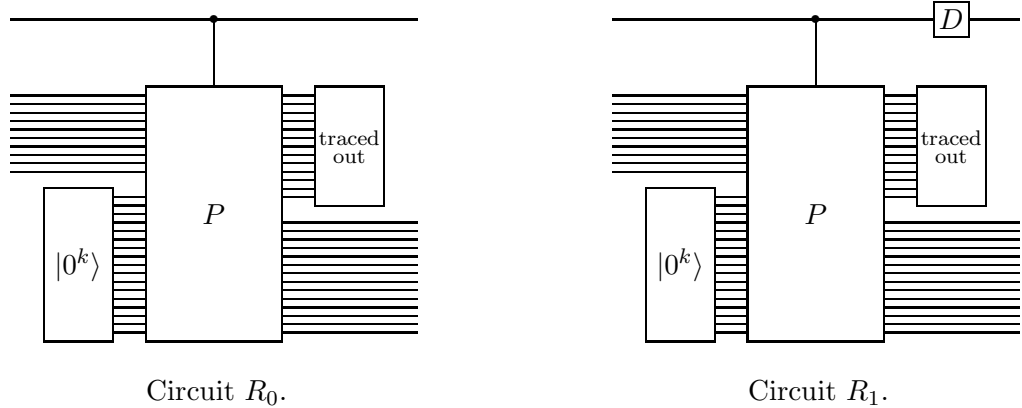
\begin{figure}[t]
  \begin{center}
    \rule{1mm}{0in}
    \begin{minipage}[t]{3.1in}
      \begin{center}
\setlength{\unitlength}{1400sp}%
\begingroup\makeatletter\ifx\SetFigFont\undefined%
\gdef\SetFigFont#1#2#3#4#5{%
  \reset@font\fontsize{#1}{#2pt}%
  \fontfamily{#3}\fontseries{#4}\fontshape{#5}%
  \selectfont}%
\fi\endgroup%
\begin{picture}(7324,6024)(1189,-6673)
\thinlines
\put(4801,-961){\circle*{150}}
\put(4801,-961){\line( 0,-1){1200}}
\put(3601,-6661){\framebox(2400,4500){$P$}}
\put(1801,-6661){\framebox(1200,2700){$\ket{0^k}$}}
\put(6601,-4261){\framebox(1200,2100)[c]{\parbox{0.6in}{\scriptsize
\centering{traced\\[-1mm] out}}}}
\put(6901,-961){\line( 1, 0){600}}
\put(1201,-2311){\line( 1, 0){2400}}
\put(1201,-2461){\line( 1, 0){2400}}
\put(1201,-2611){\line( 1, 0){2400}}
\put(1201,-2761){\line( 1, 0){2400}}
\put(1201,-2911){\line( 1, 0){2400}}
\put(1201,-3061){\line( 1, 0){2400}}
\put(1201,-3211){\line( 1, 0){2400}}
\put(1201,-3361){\line( 1, 0){2400}}
\put(1201,-3511){\line( 1, 0){2400}}
\put(1201,-3661){\line( 1, 0){2400}}
\put(1201,-961){\line( 1, 0){5700}}
\put(7501,-961){\line( 1, 0){900}}
\put(3001,-4111){\line( 1, 0){600}}
\put(3001,-4261){\line( 1, 0){600}}
\put(3001,-4411){\line( 1, 0){600}}
\put(3001,-4561){\line( 1, 0){600}}
\put(3001,-4711){\line( 1, 0){600}}
\put(3001,-4861){\line( 1, 0){600}}
\put(3001,-5011){\line( 1, 0){600}}
\put(3001,-5161){\line( 1, 0){600}}
\put(3001,-5311){\line( 1, 0){600}}
\put(3001,-5461){\line( 1, 0){600}}
\put(3001,-5611){\line( 1, 0){600}}
\put(3001,-5761){\line( 1, 0){600}}
\put(3001,-5911){\line( 1, 0){600}}
\put(3001,-6061){\line( 1, 0){600}}
\put(3001,-6211){\line( 1, 0){600}}
\put(3001,-6361){\line( 1, 0){600}}
\put(3001,-6511){\line( 1, 0){600}}
\put(6001,-2311){\line( 1, 0){600}}
\put(6001,-2461){\line( 1, 0){600}}
\put(6001,-2611){\line( 1, 0){600}}
\put(6001,-2761){\line( 1, 0){600}}
\put(6001,-2911){\line( 1, 0){600}}
\put(6001,-3061){\line( 1, 0){600}}
\put(6001,-3361){\line( 1, 0){600}}
\put(6001,-3211){\line( 1, 0){600}}
\put(6001,-3511){\line( 1, 0){600}}
\put(6001,-3661){\line( 1, 0){600}}
\put(6001,-3811){\line( 1, 0){600}}
\put(6001,-3961){\line( 1, 0){600}}
\put(6001,-4111){\line( 1, 0){600}}
\put(6001,-4561){\line( 1, 0){2400}}
\put(6001,-4711){\line( 1, 0){2400}}
\put(6001,-4861){\line( 1, 0){2400}}
\put(6001,-5011){\line( 1, 0){2400}}
\put(6001,-5161){\line( 1, 0){2400}}
\put(6001,-5311){\line( 1, 0){2400}}
\put(6001,-5461){\line( 1, 0){2400}}
\put(6001,-5611){\line( 1, 0){2400}}
\put(6001,-5761){\line( 1, 0){2400}}
\put(6001,-5911){\line( 1, 0){2400}}
\put(6001,-6061){\line( 1, 0){2400}}
\put(6001,-6211){\line( 1, 0){2400}}
\put(6001,-6361){\line( 1, 0){2400}}
\put(6001,-6511){\line( 1, 0){2400}}
\end{picture}
\\[3mm]
Circuit $R_0$.
      \end{center}
    \end{minipage}
    \begin{minipage}[t]{3.1in}
      \begin{center}
\setlength{\unitlength}{1400sp}%
\begingroup\makeatletter\ifx\SetFigFont\undefined%
\gdef\SetFigFont#1#2#3#4#5{%
  \reset@font\fontsize{#1}{#2pt}%
  \fontfamily{#3}\fontseries{#4}\fontshape{#5}%
  \selectfont}%
\fi\endgroup%
\begin{picture}(7324,6024)(1189,-6673)
\thinlines
\put(4801,-961){\circle*{150}}
\put(4801,-961){\line( 0,-1){1200}}
\put(3601,-6661){\framebox(2400,4500){$P$}}
\put(1801,-6661){\framebox(1200,2700){$\ket{0^k}$}}
\put(6601,-4261){\framebox(1200,2100)[c]{\parbox{0.6in}{\scriptsize
\centering{traced\\[-1mm] out}}}}
\put(6901,-1261){\framebox(600,600){$D$}}
\put(1201,-2311){\line( 1, 0){2400}}
\put(1201,-2461){\line( 1, 0){2400}}
\put(1201,-2611){\line( 1, 0){2400}}
\put(1201,-2761){\line( 1, 0){2400}}
\put(1201,-2911){\line( 1, 0){2400}}
\put(1201,-3061){\line( 1, 0){2400}}
\put(1201,-3211){\line( 1, 0){2400}}
\put(1201,-3361){\line( 1, 0){2400}}
\put(1201,-3511){\line( 1, 0){2400}}
\put(1201,-3661){\line( 1, 0){2400}}
\put(1201,-961){\line( 1, 0){5700}}
\put(7501,-961){\line( 1, 0){900}}
\put(3001,-4111){\line( 1, 0){600}}
\put(3001,-4261){\line( 1, 0){600}}
\put(3001,-4411){\line( 1, 0){600}}
\put(3001,-4561){\line( 1, 0){600}}
\put(3001,-4711){\line( 1, 0){600}}
\put(3001,-4861){\line( 1, 0){600}}
\put(3001,-5011){\line( 1, 0){600}}
\put(3001,-5161){\line( 1, 0){600}}
\put(3001,-5311){\line( 1, 0){600}}
\put(3001,-5461){\line( 1, 0){600}}
\put(3001,-5611){\line( 1, 0){600}}
\put(3001,-5761){\line( 1, 0){600}}
\put(3001,-5911){\line( 1, 0){600}}
\put(3001,-6061){\line( 1, 0){600}}
\put(3001,-6211){\line( 1, 0){600}}
\put(3001,-6361){\line( 1, 0){600}}
\put(3001,-6511){\line( 1, 0){600}}
\put(6001,-2311){\line( 1, 0){600}}
\put(6001,-2461){\line( 1, 0){600}}
\put(6001,-2611){\line( 1, 0){600}}
\put(6001,-2761){\line( 1, 0){600}}
\put(6001,-2911){\line( 1, 0){600}}
\put(6001,-3061){\line( 1, 0){600}}
\put(6001,-3361){\line( 1, 0){600}}
\put(6001,-3211){\line( 1, 0){600}}
\put(6001,-3511){\line( 1, 0){600}}
\put(6001,-3661){\line( 1, 0){600}}
\put(6001,-3811){\line( 1, 0){600}}
\put(6001,-3961){\line( 1, 0){600}}
\put(6001,-4111){\line( 1, 0){600}}
\put(6001,-4561){\line( 1, 0){2400}}
\put(6001,-4711){\line( 1, 0){2400}}
\put(6001,-4861){\line( 1, 0){2400}}
\put(6001,-5011){\line( 1, 0){2400}}
\put(6001,-5161){\line( 1, 0){2400}}
\put(6001,-5311){\line( 1, 0){2400}}
\put(6001,-5461){\line( 1, 0){2400}}
\put(6001,-5611){\line( 1, 0){2400}}
\put(6001,-5761){\line( 1, 0){2400}}
\put(6001,-5911){\line( 1, 0){2400}}
\put(6001,-6061){\line( 1, 0){2400}}
\put(6001,-6211){\line( 1, 0){2400}}
\put(6001,-6361){\line( 1, 0){2400}}
\put(6001,-6511){\line( 1, 0){2400}}
\end{picture}
\\[3mm]
    Circuit $R_1$.
      \end{center}
    \end{minipage}
  \end{center}
  \caption{Circuits output by the reduction.}
  \label{fig:R0_and_R1}
\end{figure}

Finally, let $R_0$ and $R_1$ be circuits constructed from $P$ and $D$ as
described in Figure~\ref{fig:R0_and_R1}.
Here, the input qubits to $R_0$ and $R_1$ correspond to the input qubits
of $Q_0$ or $Q_1$, which $P$ simulates, as well as the control qubit of $P$.
The remaining $k$ qubits are initialized to the zero state, which
is required for the correct functioning of $P$.
The qubits that are output by $P$ include the control qubit, the $m$ qubits
representing the output of $Q_0$ or $Q_1$, and the $l$ ``garbage'' qubits that
are traced out when simulating $Q_0$ or $Q_1$.
The circuits $R_0$ and $R_1$, however, reverse the roles of the
output qubits and garbage qubits of $P$.
Specifically, the garbage qubits of $P$ together with the control qubit
are the output qubits of $R_0$ and $R_1$, while the qubits of $P$ corresponding
to the output of $Q_0$ or $Q_1$ are traced out by $R_0$ and $R_1$.
It is this reversal that is the key to the reduction.
The circuits $R_0$ and $R_1$ differ only in that $R_1$ includes the
decoherence gate on the control qubit after $P$ is performed while $R_0$
does not.

When either of the circuits $R_0$ and $R_1$ is given an input in which the
control qubit is in a superposition of state 0 and 1, possibly entangled with
the other input qubits, in effect both of the circuits $Q_0$ and $Q_1$ are run.
The idea of the reduction is that if the outputs of $Q_0$ and $Q_1$ are close
on their respective inputs, then when run in superposition discarding these
outputs will not destroy the coherence of the control qubit, and thus the
outputs of $R_0$ and $R_1$ will differ significantly because of the action of
the decoherence gate.
If the outputs of $Q_0$ and $Q_1$ are distinguishable, however,
discarding the output qubits of $Q_0$ or $Q_1$ is tantamount to
decoherence of the control qubit, and so there is no significant
difference between $R_0$ and $R_1$ in this case as the decoherence
gate is effectively redundant.

Formalizing this argument and using suitable parameters allows us to
conclude that $\prob{QCD}_{1,1/4}$ is \class{QIP}-hard.
Extending hardness to $\prob{QCD}_{2-\varepsilon,\varepsilon}$
can be accomplished by using a variant of Sahai and Vadhan's
method of ``polarizing'' samplable distributions \cite{SahaiV03}
applied to admissible transformations.


\subsection{Proof of Theorem~\ref{theorem:main}}

This section contains a more formal proof that
$\prob{QCD}_{2-\varepsilon,\varepsilon}$ is
\class{QIP}-hard for any constant $\varepsilon >0$, as described in the
previous subsection.
As $\prob{QCD}_{2-\varepsilon,\varepsilon}\in\class{QIP}$, this will
imply Theorem~\ref{theorem:main}.

Let $Q_0$ and $Q_1$ be mixed-state circuits of type $(n,m)$, and consider
the circuit construction described in Section~\ref{sec:proof-overview}.
To be more precise, let $\mathcal{H} = \mathcal{H}(\Sigma^n)$ denote the
space corresponding to the input qubits of $Q_0$ and $Q_1$ and let
$\mathcal{K} = \mathcal{K}(\Sigma^m)$ denote the space corresponding to the
output qubits of $Q_0$ and $Q_1$.
As discussed in Section~\ref{sec:circuits}, it is possible to efficiently
construct unitary circuits $P_0$ and $P_1$, acting on $n+k=m+l$
qubits for some choice of $k$ and $l$, that simulate $Q_0$ and $Q_1$.
Specifically, if $\mathcal{E} = \mathcal{E}(\Sigma^k)$ and
$\mathcal{F} = \mathcal{F}(\Sigma^l)$, then $P_0$ and $P_1$
induce unitary transformations
$U_0,U_1\in\mathbf{U}(\mathcal{H}\otimes\mathcal{E},\mathcal{K}\otimes
\mathcal{F})$ satisfying $Q_i(\rho) = \op{tr}_{\mathcal{F}}
U_i(\rho\otimes\ket{0^k}\bra{0^k})U_i^{\dagger}$ for $i=0,1$.

Next, let $\mathcal{A} = \mathcal{A}(\Sigma)$ be the space corresponding to
a single qubit, and define a unitary operator
$U\in\mathbf{U}(\mathcal{A}\otimes\mathcal{H}\otimes\mathcal{E},
\mathcal{A}\otimes\mathcal{K}\otimes\mathcal{F})$ by the the equations
$U(\ket{0}\ket{\psi}) = \ket{0}\, U_0 \ket{\psi}$ and
$U(\ket{1}\ket{\psi}) = \ket{1}\, U_1 \ket{\psi}$
for every $\ket{\psi}\in\mathcal{H}\otimes\mathcal{E}$.
It is possible to construct a unitary circuit $P$ whose operation is
described by $U$ that has size polynomial in the sizes of $P_0$ and $P_1$.
Specifically, this may be done by replacing each gate of $P_0$ and $P_1$
by a similar gate that is appropriately controlled by the qubit corresponding
to the space $\mathcal{A}$ and running the two circuits one after the other.
The controlled gates are of constant size and may either be implemented
directly or approximated with very high accuracy depending on the basis gates
being considered.
See Nielsen and Chuang~\cite[section 4.3]{NielsenC00} for further information
on such constructions.
We can assume without loss of generality that $P$ acts on exactly those qubits
$P_0$ and $P_1$ act on plus the control qubit; any ancilla required by
$P$ can be included in $P_0$ and $P_1$.

Finally, the circuits $R_0$ and $R_1$ described in Figure~\ref{fig:R0_and_R1}
correspond to admissible operations in the set
$\mathbf{T}(\mathcal{A}\otimes\mathcal{H},\mathcal{A}\otimes\mathcal{F})$,
and can be described more precisely by
\[
R_0(X) = \op{tr}_{\mathcal{K}}\left(P\left(X\otimes\ket{0^k}\bra{0^k}
\right)\right),\;\;\;\;\;
R_1(X) = (D\otimes I_{\mathbf{L}(\mathcal{F})})
\left(\op{tr}_{\mathcal{K}}\left(P\left(X\otimes\ket{0^k}\bra{0^k}\right)
\right)\right)
\]
for every $X\in\mathbf{L}(\mathcal{A}\otimes\mathcal{H})$.
Here, the decoherence operation $D$ is acting on $\mathcal{A}$, i.e.,
$D\in\mathbf{T}(\mathcal{A},\mathcal{A})$.
The space $\mathcal{K}$, which corresponds to the output qubits
of $Q_0$ and $Q_1$, is the space that is traced out by $R_0$ and $R_1$, while
the output qubits of $R_0$ and $R_1$ consist of the control qubit and the
``garbage'' qubits of $P_0$ and $P_1$, which correspond to $\mathcal{F}$.
Descriptions of these two new circuits can be computed in polynomial time
given descriptions of $Q_0$ and $Q_1$.

The following lemma formalizes the intuition discussed previously that $R_0$
and $R_1$ act very differently if $Q_0$ and $Q_1$ can be made to have outputs
that have high fidelity with one another.

\begin{lemma}\label{lemma:reduction}
  $\dnorm{R_0 - R_1} = \max\left\{F(Q_0(\rho_0),Q_1(\rho_1))
  \,:\,\rho_0,\rho_1\in\mathbf{D}(\mathcal{H})\right\}$.
\begin{proof}
Let $\rho_0,\rho_1\in\mathbf{D}(\mathcal{H})$ be any two states.
We will show that $\dnorm{R_0 - R_1} \geq F(Q_0(\rho_0),Q_1(\rho_1))$.
Define $W_0,W_1\in\mathbf{L}(\mathcal{H},\mathcal{K}\otimes\mathcal{F})$ as
$W_i = U_i(I_{\mathcal{H}}\otimes \ket{0^k})$ for $i=0,1$, where $U_i$
is the unitary operator corresponding to circuit $P_i$.
Each $W_i$ is a unitary embedding that effectively concatenates $k$ ancilla
qubits to a vector in $\mathcal{H}$, and then performs $U_i$ on the
resulting vector.
Let $\ket{\psi_0},\ket{\psi_1}\in\mathcal{H}\otimes\mathcal{G}$ be
any purifications of $\rho_0,\rho_1$, respectively, where $\mathcal{G}$
is any Hilbert space large enough to admit such purifications.
Let $\ket{\psi} = \frac{1}{\sqrt{2}}\ket{0}\ket{\psi_0} +
\frac{1}{\sqrt{2}}\ket{1}\ket{\psi_1}$
and consider the action of $R_0$ and $R_1$ on $\ket{\psi}\bra{\psi}$ (where
the circuits act trivially on the space $\mathcal{G}$).
The circuits are identical aside from the decoherence gate.
Immediately after
the circuit $P$ is performed but before the qubits corresponding to the space
$\mathcal{K}$ are traced out, the state obtained for both circuits will be
$\ket{\phi}\bra{\phi}$, for
$\ket{\phi} = \frac{1}{\sqrt{2}}\ket{0}\ket{\phi_0} +
\frac{1}{\sqrt{2}}\ket{1}\ket{\phi_1}$, where
$\ket{\phi_0} = (W_0\otimes I_{\mathcal{G}})\ket{\psi_0}$ and
$\ket{\phi_1} = (W_1\otimes I_{\mathcal{G}})\ket{\psi_1}$.
The output of circuit $R_0$ can therefore be written as
\[
\frac{1}{2}\op{tr}_{\mathcal{K}}\left(
\ket{0}\bra{0}\otimes \ket{\phi_0}\bra{\phi_0}
\,+\,\ket{0}\bra{1}\otimes \ket{\phi_0}\bra{\phi_1}
\,+\,\ket{1}\bra{0}\otimes \ket{\phi_1}\bra{\phi_0}
\,+\,\ket{1}\bra{1}\otimes \ket{\phi_1}\bra{\phi_1}
\right)
\]
while the output of circuit $R_1$ is
\[
\frac{1}{2}\op{tr}_{\mathcal{K}}\left(
\ket{0}\bra{0}\otimes \ket{\phi_0}\bra{\phi_0}
\,+\,\ket{1}\bra{1}\otimes \ket{\phi_1}\bra{\phi_1}
\right).
\]
This is because the effect of the decoherence gate is to eliminate
the cross-terms $\ket{0}\bra{1}\otimes\ket{\phi_0}\bra{\phi_1}$
and $\ket{1}\bra{0}\otimes\ket{\phi_1}\bra{\phi_0}$.
As
$\ket{\phi_0},\ket{\phi_1}\in\mathcal{K}\otimes\mathcal{F}\otimes\mathcal{G}$
are purifications of $Q_0(\rho_0)$ and $Q_1(\rho_1)$, respectively, we may
conclude by Theorem~\ref{theorem:diamnorm-tracenorm} and
Lemma~\ref{lemma:fidelity-tracenorm} that
\begin{multline*}
\dnorm{R_0 - R_1}  \geq
\left\| (R_0\otimes I_{\mathbf{L}(\mathcal{G})})(\ket{\phi}\bra{\phi})
-(R_1\otimes I_{\mathbf{L}(\mathcal{G})})(\ket{\phi}\bra{\phi})
\right\|_{\mathrm{tr}}\\
=
\frac{1}{2}\left\|
\ket{0}\bra{1}\otimes\op{tr}_{\mathcal{K}}\ket{\phi_0}\bra{\phi_1} +
\ket{1}\bra{0}\otimes\op{tr}_{\mathcal{K}}\ket{\phi_1}\bra{\phi_0}
\right\|_{\mathrm{tr}}
=
\left\|
\op{tr}_{\mathcal{K}}\ket{\phi_0}\bra{\phi_1}
\right\|_{\mathrm{tr}}
= F(Q_0(\rho_0),Q_1(\rho_1)).
\end{multline*}

Next, by Lemma~\ref{lemma:max-for-dnorm} we have
$\dnorm{R_0 - R_1} = \left\|
(R_0\otimes I_{\mathbf{L}(\mathcal{G})})(\ket{\psi}\bra{\psi}) -
(R_1\otimes I_{\mathbf{L}(\mathcal{G})})(\ket{\psi}\bra{\psi})
\right\|_{\mathrm{tr}}$
for some Hilbert space $\mathcal{G}$ and unit vector
$\ket{\psi}\in\mathcal{A}\otimes\mathcal{H}\otimes\mathcal{G}$.
As $\ket{\psi}$ is a unit vector we may write
$\ket{\psi} = \sqrt{p}\,\ket{0}\ket{\psi_0} + \sqrt{1-p}\,\ket{1}\ket{\psi_1}$
for $\ket{\psi_0},\ket{\psi_1}\in\mathcal{H}\otimes\mathcal{G}$ unit vectors
and $p\in [0,1]$.
Let $\ket{\phi_i}=(W_i\otimes I_{\mathcal{G}})\ket{\psi_i}$
and $\rho_i = \op{tr}_{\mathcal{G}}\ket{\psi_i}\bra{\psi_i}$, for $i=0,1$.
We have
\begin{eqnarray*}
\lefteqn{
\left\|(R_0\otimes I_{\mathbf{L}(\mathcal{G})})(\ket{\psi}\bra{\psi}) -
(R_1\otimes I_{\mathbf{L}(\mathcal{G})})(\ket{\psi}\bra{\psi})
\right\|_{\mathrm{tr}}}\hspace*{2cm}\\
& = &
\sqrt{p(1-p)}\left\|
\ket{0}\bra{1}\otimes\op{tr}_{\mathcal{K}}\ket{\phi_0}\bra{\phi_1} +
\ket{1}\bra{0}\otimes\op{tr}_{\mathcal{K}}\ket{\phi_1}\bra{\phi_0}
\right\|_{\mathrm{tr}}\\
& = &
2\sqrt{p(1-p)}\left\|\op{tr}_{\mathcal{K}}\ket{\phi_0}\bra{\phi_1}
\right\|_{\mathrm{tr}}\\
& \leq & F(Q_0(\rho_0),Q_1(\rho_1)).
\end{eqnarray*}
This completes the proof of the lemma.
\end{proof}
\end{lemma}

This lemma and the above construction imply that
$\prob{CI}_{a,b}\leq_m^p \prob{QCD}_{a,b}$ for all $a,b\in[0,1]$ with $b < a$.
As $\prob{CI}_{1,1/4}$ is a complete promise problem for \class{QIP}
and $\prob{QCD}_{1,1/4}$ is in \class{QIP}, we have that
$\prob{QCD}_{1,1/4}$ is \class{QIP}-complete.

Finally, we can extend the \class{QIP}-hardness of $\prob{QCD}_{1,1/4}$
to instances of the Quantum Circuit Distinguishability problem with a much
stronger promise.
This fact is based on a generalization of the ``polarization'' method developed
by Sahai and Vadhan \cite{SahaiV03} in the context of statistical
zero-knowledge.

\begin{theorem}\label{theorem:polarize}
Let $a,b\in(0,2)$ satisfy $2 b < a^2$.
There exists a deterministic, polynomial-time procedure that, when given as
input $(R_0, R_1, 1^n)$, where $R_0$ and $R_1$ are mixed-state quantum
circuits, outputs quantum circuits $(S_0, S_1)$ such that
\begin{mylist}{5mm}
\item[1.]
$\dnorm{R_0 - R_1} \leq  b \;\Rightarrow\; \dnorm{S_0 - S_1} <  2^{-n}$,
and
\item[2.]
$\dnorm{R_0 - R_1} \geq  a  \;\Rightarrow\; \dnorm{S_0 - S_1} >  2 - 2^{-n}$.
\end{mylist}
\end{theorem}

\noindent
Sahai and Vadhan proved this theorem for polynomial-time samplable
distributions, and it was observed in Ref.~\cite{Watrous02} that the theorem
carries over to polynomial-time preparable quantum states.
In the present case, the theorem must be extended to admissible operations.

\begin{lemma}\label{lemma_direct_product}
  If $\Phi_1, \Phi_2 \in \transform{H,K}$ satisfy 
  $\dnorm{\Phi_1 - \Phi_2} = \varepsilon$, then
  \[
  2 - 2e^{ \frac{-k \varepsilon^2}{8} }
  < \dnorm{ \Phi_1^{\tensor k} - \Phi_2^{\tensor k}}
  \leq  k \varepsilon.
  \]
  \begin{proof}
    Let $\mathcal{F}$ be a Hilbert space of dimension equal to that of
    $\mathcal{H}$, and let $Y\in\mathbf{L}(\mathcal{H}\otimes\mathcal{F})$
    satisfy $\tnorm{Y} = 1$ and
    \[
    \tnorm{(\Phi_1\otimes I_{\mathbf{L}(\mathcal{F})})(Y) -
      (\Phi_2\otimes I_{\mathbf{L}(\mathcal{F})})(Y)} = \dnorm{\Phi_1-\Phi_2}
    = \varepsilon.
    \]
    Then because $\tnorm{Y^{\otimes k}} = 1$ we have
    \begin{align*}
      & \hspace{-4mm} \dnorm{ \Phi_1^{\tensor k} - \Phi_2^{\tensor k}}\\
      & =
      \max\left\{
      \tnorm{(\Phi_1\tensor I_{\mathbf{L}(\mathcal{F})})^{\tensor k}(X)
      - (\Phi_2\tensor I_{\mathbf{L}(\mathcal{F})})^{\tensor k}(X)}\,:\,
      X\in\mathbf{L}((\mathcal{H}\otimes\mathcal{F})^{\otimes k}),
      \,\tnorm{X} = 1\right\}\\
      & \geq 
      \tnorm{\left((\Phi_1 \tensor I_{\mathbf{L}(\mathcal{F})})(Y)
    \right)^{\tensor k}
    - \left((\Phi_1 \tensor I_{\mathbf{L}(\mathcal{F})})(Y)
    \right)^{\tensor k}}\\
      & \geq  2 - 2e^{ \frac{-k \varepsilon^2}{8}}.
    \end{align*}
    The last inequality follows from the result for states analogous to
    what is here being proved \cite{Watrous02}.

    The second inequality will be proved by induction.
    The base case $k=1$ is trivial:
    \[
    \dnorm{ \Phi_1^{\tensor 1} - \Phi_2^{\tensor 1} }
    = \dnorm{ \Phi_1 - \Phi_2 } = \varepsilon.
    \]
    Assume then that $k>1$,
    and define $\Psi_i = \Phi_i^{\tensor{(k-1)}}$ for $i\in\{1,2\}$.
    We have
    \begin{eqnarray*}
      \dnorm{ \Phi_1^{\tensor k} - \Phi_2^{\tensor k}}
      & = & \dnorm{ \Psi_1 \tensor \Phi_1 - \Psi_2 \tensor \Phi_2 } \\
      & = & \dnorm{ \Psi_1 \tensor \Phi_1 - \Psi_2 \tensor \Phi_1 +
	\Psi_2 \tensor \Phi_1 - \Psi_2 \tensor \Phi_2 } \\
      & \leq & \dnorm{(\Psi_1 - \Psi_2) \tensor \Phi_1 } +
      \dnorm{\Psi_2 \tensor (\Phi_1 - \Phi_2) } \\
      & = & \dnorm{\Psi_1-\Psi_2} \dnorm{\Phi_1} + \dnorm{\Psi_2} 
      \dnorm{\Phi_1 - \Phi_2}.
    \end{eqnarray*}
    Because the diamond norm of any admissible transformation is one (see
    \cite{AharonovK+98} for a proof), we obtain
    \[
    \dnorm{ \Psi_1 - \Psi_2 } \dnorm{\Phi_1} + \dnorm{\Psi_2}
    \dnorm{\Phi_1 - \Phi_2}
    \leq (k - 1) \varepsilon + \varepsilon = k \varepsilon
    \]
    as required.
  \end{proof}
\end{lemma}

\begin{lemma}\label{lemma_shrink}
  There is a deterministic polynomial-time procedure that, on input
  $(Q_0, Q_1, 1^r)$, where $Q_0, Q_1$ are descriptions of mixed-state quantum
  circuits, produces as output descriptions of two quantum circuits,
  $(R_0, R_1)$ satisfying
  \[
  2 - 2 \exp\left( - \frac{r}{8} \dnorm{Q_0 - Q_1}^2 \right)
  \leq  \dnorm{R_0 - R_1}
  \leq r \dnorm{Q_0 - Q_1}.
  \]
  \begin{proof}
    For $i=0,1$, construct $R_i$ by placing $r$ copies of the circuit
    $Q_i$ in parallel.  Then $R_i = Q_i^{\tensor r}$, and the bounds
    on $\dnorm{R_0 - R_1}$ follow from Lemma~\ref{lemma_direct_product}.
  \end{proof}
\end{lemma}

\begin{prop}\label{prop_dnorm_split}
  Let $\Phi_0, \Phi_1 \in \transform{H,K}$ and 
  $\Psi_0, \Psi_1 \in \transform{F,G}$.
  Define
  \begin{eqnarray*}
    \Xi_0 & = & \frac{1}{2} \Phi_0 \tensor \Psi_0 + 
    \frac{1}{2} \Phi_1 \tensor \Psi_1, \\
    \Xi_1 & = & \frac{1}{2} \Phi_0 \tensor \Psi_1 + 
    \frac{1}{2} \Phi_0 \tensor \Psi_1.
  \end{eqnarray*}
  Then
  $\dnorm{\Xi_0-\Xi_1} = \frac{1}{2}\dnorm{\Phi_0-\Phi_1} \cdot 
  \dnorm{\Psi_0 - \Psi_1}$.

  \begin{proof}
    Using $\Xi_0, \Xi_1$ as in the proposition, we have
    \begin{eqnarray*}
      \dnorm{\Xi_0 - \Xi_1}
      & = & \dnorm{ \frac{1}{2} \Phi_0 \tensor \Psi_0 + 
	\frac{1}{2} \Phi_1 \tensor \Psi_1
	- \frac{1}{2} \Phi_0 \tensor \Psi_1 
	- \frac{1}{2} \Phi_0 \tensor \Psi_1} \\
      & = & \dnorm{ \frac{1}{2} (\Phi_0 - \Phi_1) 
	\tensor (\Psi_0 - \Psi_1) } \\
      & = & \frac{1}{2} \dnorm{ \Phi_0 - \Phi_1 } \cdot 
      \dnorm{\Psi_0 - \Psi_1}.
    \end{eqnarray*}
    as desired.
  \end{proof}
\end{prop}

\begin{lemma}\label{lemma_grow}
  There is a deterministic polynomial-time procedure that,
  on input $(Q_0, Q_1, 1^r)$, where $Q_0, Q_1$ are descriptions of
  mixed-state quantum circuits, produces as output descriptions of two
  quantum circuits $(R_0, R_1)$ satisfying
  \[
  \dnorm{R_0 - R_1} = 2 \left( \frac{\dnorm{Q_0 - Q_1}}{2} \right)^r.
  \]
  \begin{proof}
    The circuit $R_0$ performs a transformation defined as
    \[
    R_0 =
    \frac{1}{2^{r-1}}\sum_{\stackrel{\scriptstyle x_1,\ldots,x_r\in\{0,1\}}
      {\scriptstyle x_1 + \cdots + x_r \equiv 0\,(\op{mod} 2)}}
    Q_{x_1}\otimes\cdots\otimes Q_{x_r}
    \]
    while $R_1$ performs a similar transformation defined as
    \[
    R_1 =
    \frac{1}{2^{r-1}}\sum_{\stackrel{\scriptstyle x_1,\ldots,x_r\in\{0,1\}}
      {\scriptstyle x_1 + \cdots + x_r \equiv 1\,(\op{mod} 2)}}
    Q_{x_1}\otimes\cdots\otimes Q_{x_r}.
    \]
    These circuits are effectively running $r$ copies of $Q_0$ and/or $Q_1$
    in parallel, with the choice of $Q_0$ or $Q_1$ determined uniformly at
    random subject to the constraint that $R_0$ applies an even number
    of copies of $Q_1$ while $R_1$ applies an odd number.
    Such circuits may be constructed in time polynomial in the sizes of
    $Q_0$ and $Q_1$.
    A proof by induction based on Proposition \ref{prop_dnorm_split}
    establishes that $R_0$ and $R_1$ have the required property.
  \end{proof}
\end{lemma}

\begin{proof}[Proof of Theorem~\ref{theorem:polarize}]
  First, we apply the procedure given by Lemma \ref{lemma_shrink} to
  $(Q_0, Q_1, 1^r)$, with
  \[
  r = \ceil{\log(16n) / \log( a^2 / (2 b))},
  \]
  obtaining circuits $(Q_0^\prime, Q_1^\prime)$ satisfying
  \begin{eqnarray*}
    \dnorm{Q_0 - Q_1} < b
    & \Rightarrow &
    \dnorm{Q_0^\prime - Q_1^\prime} < 2 (b/2)^r \\
    \dnorm{Q_0 - Q_1} > a
    & \Rightarrow &
    \dnorm{Q_0^\prime - Q_1^\prime} > 2 (a/2)^r
  \end{eqnarray*}
  Next, we apply the procedure given by Lemma \ref{lemma_grow} to
  $(Q_0^\prime, Q_1^\prime, 1^s)$, where $s = \floor{ (b / 2)^{-r} / 4}$,
  obtaining circuits $(Q_0^{\prime\prime}, Q_1^{\prime\prime})$ satisfying
  \begin{eqnarray*}
    \dnorm{Q_0 - Q_1} < b
    & \Rightarrow &
    \dnorm{Q_0'' - Q_1''} < 2 (b/2)^r (b/2)^{-r}/4 = 1/2 \\
    \dnorm{Q_0 - Q_1} > a
    & \Rightarrow &
    \dnorm{Q_0'' - Q_1''} >  2 - 2 \exp(-\frac{s}{2}(a/2)^{2r} )
    \geq 2 - 2 e^{-2n + 1}.
  \end{eqnarray*}
  Finally, we apply the construction of Lemma \ref{lemma_shrink}
  once more, to $(Q_0^{\prime\prime}, Q_1^{\prime\prime}, 1^t)$,
  where $t = \ceil{(n+1)/2}$, obtaining circuits $(R_0, R_1)$
  satisfying
  \begin{eqnarray*}
    \dnorm{Q_0 - Q_1} < b
    & \Rightarrow &
    \dnorm{R_0 - R_1} < (1/2)^{(n+1)/2} (1/2)^{(n - 1)/2} = 2^{-n} \\
    \dnorm{Q_0 - Q_1} > a
    & \Rightarrow &
    \dnorm{R_0 - R_1} > (2 - 2 e^{-2n + 1})^{\ceil{(n+1)/2}}
    (1/2)^{\ceil{(n+1)/2} - 1} \geq 2 - 2^{-n}.
  \end{eqnarray*}
  The circuits $(R_0, R_1)$ have size polynomial in $r, s, t$ and the size of
  the original circuits $(Q_0, Q_1)$.
  Because $r, s, t$ are bounded by polynomials in $n$, the size of the
  constructed circuits is polynomial in the size of the input.
\end{proof}

\noindent
Theorem~\ref{theorem:polarize} implies that
$\prob{QCD}_{1,1/4} \leq_m^p \prob{QCD}_{2-\varepsilon,\varepsilon}$
for every $\varepsilon >0$, which proves Theorem~\ref{theorem:main}.


\section{Conclusion}
\label{sec:conclusion}

We have demonstrated that the problem of distinguishing mixed-state
quantum circuits is a complete promise problem for the class \class{QIP}.
Because \class{QIP} contains \class{PSPACE}, we conclude that this problem is
\class{PSPACE}-hard, whereas its classical analogue is contained in the class
\class{AM} and its unitary quantum circuit analogue is in \class{QMA}.

Some open questions relating to this paper follow.
\begin{mylist}{5mm}
\item[$\bullet$]
Does the \class{QIP}-completeness of the \prob{QCD} problem shed any light on
properties of \class{QIP}?  For instance, is \class{QIP} is closed under
complementation?  Is $\prob{QCD}\in\class{PSPACE}$, which would imply
$\class{QIP} = \class{PSPACE}$?

\item[$\bullet$]
There are interesting questions and results relating to implementations
of quantum computers that deal with unitary circuits with mixed-state inputs.
(See, e.g., \cite{AmbainisS+00,KnillL98}.)
Analogues of the \prob{QCD} problem can be defined for this setting.
For example, one might consider unitary circuits that act on some
collection of inputs together with a collection of qubits in the
totally mixed state.
How hard is the \prob{QCD} problem in this context?

\item[$\bullet$]
Because it is not known whether $\class{QIP} = \class{PSPACE}$, the
\prob{QCD} problem is a candidate problem for 
$\class{QIP}\backslash\class{PSPACE}$.
Are there any reasonable non-promise problem candidates for problems in
$\class{QIP}$ but not in $\class{PSPACE}$?

\end{mylist}


\subsection*{Acknowledgments}

This research was supported by Canada's {\sc Nserc}, the Canadian Institute for
Advanced Research (CIAR), and the Canada Research Chairs Program.



\begin{thebibliography}{10}

\bibitem{AharonovK+98}
D.~Aharonov, A.~Kitaev, and N.~Nisan.
\newblock Quantum circuits with mixed states.
\newblock In {\em Proceedings of the Thirtieth Annual ACM Symposium on Theory
  of Computing}, pages 20--30, 1998.

\bibitem{AmbainisS+00}
A.~Ambainis, L.~Schulman, and U.~Vazirani.
\newblock Computing with highly mixed states.
\newblock In {\em Proceedings of the 32nd Annual Symposium on the Theory of
  Computing}, 2000.

\bibitem{Choi75}
M.-D. Choi.
\newblock Completely positive linear maps on complex matrices.
\newblock {\em Linear Algebra and its Applications}, 10(3):285--290, 1975.

\bibitem{JanzingW+03}
D.~Janzing, P.~Wocjan, and T.~Beth.
\newblock ``{I}dentity check'' is {QMA}-complete.
\newblock Available as arXiv.org e-Print \mbox{quant-ph}/0305050, 2003.

\bibitem{Kitaev97}
A.~Kitaev.
\newblock Quantum computations: algorithms and error correction.
\newblock {\em Russian Mathematical Surveys}, 52(6):1191--1249, 1997.

\bibitem{KitaevS+02}
A.~Kitaev, A.~Shen, and M.~Vyalyi.
\newblock {\em Classical and Quantum Computation}, volume~47 of {\em Graduate
  Studies in Mathematics}.
\newblock American Mathematical Society, 2002.

\bibitem{KitaevW00}
A.~Kitaev and J.~Watrous.
\newblock Parallelization, amplification, and exponential time simulation of
  quantum interactive proof system.
\newblock In {\em Proceedings of the 32nd ACM Symposium on Theory of
  Computing}, pages 608--617, 2000.

\bibitem{KnillL98}
E.~Knill and R.~Laflamme.
\newblock On the power of one bit of quantum information.
\newblock {\em Physical Review Letters}, 81:5672--5675, 1998.

\bibitem{NielsenC00}
M.~A. Nielsen and I.~L. Chuang.
\newblock {\em Quantum Computation and Quantum Information}.
\newblock Cambridge University Press, 2000.

\bibitem{SahaiV03}
A.~Sahai and S.~Vadhan.
\newblock A complete promise problem for statistical zero-knowledge.
\newblock {\em Journal of the ACM}, 50(2):196--249, 2003.


\bibitem{Watrous02}
J.~Watrous.
\newblock Limits on the power of quantum statistical zero-knowledge.
\newblock In {\em Proceedings of the 43rd Annual Symposium on Foundations of
  Computer Science}, pages 459--468, 2002.
\newblock Full version available at
  http://www.cpsc.ucalgary.ca/\raisebox{1mm}{\tiny $\sim$}jwatrous/papers.html.

\bibitem{Watrous03-pspace}
J.~Watrous.
\newblock {PSPACE} has constant-round quantum interactive proof systems.
\newblock {\em Theoretical Computer Science}, 292(3):575--588, 2003.

\end{thebibliography}
\end{document}